\documentclass{ws-ijmpd}
\usepackage[super,compress]{cite}
\usepackage{color}
\usepackage{amssymb}
\usepackage{amsmath}
\usepackage{graphicx}
\usepackage[english]{babel}
\usepackage[T1]{fontenc}
\usepackage{hyperref}
\newcommand{\mawave}{{\mathcal G}} %
\def\eslash{\not\!\! E}
\def\gev{\; \hbox{GeV}}
\begin{document}
\title{Perspectives for an  Elko Phenomenology  using Monojets at the 14 TeV LHC}

\author{Alexandre Alves}
\address{Departamento de Ci\^encias Exatas e da Terra, Universidade Federal de S\~ao Paulo\\
Diadema - SP - Brazil\\
aalves@unifesp.br}

\author{M. Dias}
\address{Departamento de Ci\^encias Exatas e da Terra, Universidade Federal de S\~ao Paulo\\
Diadema - SP - Brazil\\
marco.dias@unifesp.br}

\author{F. de Campos}
\address{Departamento de Engenharia Ambiental, Universidade Estadual Paulista, \\ S\~ao Jos\'e dos Campos-SP-Brazil\\
fernando.carvalho@ict.unesp.br}
\maketitle

\begin{abstract}
The aim of this work is to explore the possibility to discover a fermionic field with mass dimension one, the Elko field, in the 14 TeV Large Hadron Collider (LHC), in processes with missing energy and one jet.  We explore the possibility of a triple coupling with the Higgs field, generating also a contribution to the Elko mass term, and suggest some possibilities for future studies in this field.
\end{abstract}

\ccode{PACS numbers:13.85.Rm,12.38.Bx,95.35.+d}

\noindent

\section{Introduction}
Experimental results from colliders have been used for studying unsolved problems in astrophysics, especially those which are related to dark matter (DM). The detection of the Higgs boson and the study of its decay modes can help to find the answer to this question when we consider the Higgs boson couplings in theories beyond the Standard Model (SM). These new theories might rely on enlarged symmetry groups or additional matter content, for example, suggesting new signals in colliders. A possible new signal is the dark matter production through couplings to the SM Higgs boson. Actually, this is a type of process widely studied in the called {\it Higgs portal} scenarios~\cite{portal}. Amongst the candidates for dark matter, we consider in this work the Elko spinor field \cite{elko,elko2}, an SM singlet that couples only to the Higgs boson in the framework of a Higgs portal. The Elko spinor field is a set of four eigenspinors of the charge conjugation operator. Another interesting feature is that Elko spinor fields do not belong to a standard Wigner class \cite{WIG}, breaking Lorentz symmetry (in a subtle way) by choosing a preferred direction of propagation\cite{ALS}. 

Elko dark matter can be searched for at colliders in processes with large missing energy accompanied by jets, photons, and charged leptons. In~\cite{nosso},  a search strategy for Elkos at the 14 TeV LHC was performed in the mono-$Z$ channel, with promising results. In this work we have investigated the prospects for Elko dark matter production in the monojet channel at the 14 TeV LHC in a high luminosity regime and taking into account the most relevant backgrounds.

This paper is organized as follows: in the next Section we calculate the cross-section for Elko production at the LHC, analyzing the case of two Elkos and one jet. Unlike the case studied in \cite{nosso}, we have considered a mass term for the Elko also shifted by a non-zero vacuum expected value (vev) for the Higgs scalar. In Section III we present our conclusions.

\section{Elko Signal and Phenomenological Analysis}
In this study we have considered the signal processes of an Elko pair $\stackrel{\neg}{\lambda}^S\lambda^S$ plus one jet. The Elko mass is $m_\varepsilon$ before the Higgs acquires a vev. In order to obtain the coupling of an Elko field with the Higgs boson, we look for the Lagrangian density given by~\cite{elko}
\begin{equation}
  \mathcal{L} = \partial^\mu {\stackrel{\neg}\lambda}(x)\,\partial_\mu\lambda(x)-m_\varepsilon^2\stackrel{\neg}\lambda(x)\,
  \lambda(x) +\lambda_E \lambda(x)\stackrel{\neg}\lambda(x)\phi(x)^2,
\end{equation}
identifying $\phi$ as a scalar field that can be shifted by a vev, $\langle \phi\rangle$ as 
\begin{equation}
\phi=\frac{1}{\sqrt{2}}\left(H+\langle \phi\rangle\right),
\end{equation}
 and we obtain
\begin{eqnarray}
  \mathcal{L} &=& \partial^\mu {\stackrel{\neg}\lambda}(x) \,\partial_\mu\lambda(x)-\left(m_\varepsilon^2-\lambda_E\frac{\langle \phi\rangle^2}{2} \right)\stackrel{\neg}\lambda(x)\,
  \lambda(x)+\frac{\lambda_E}{2}\lambda(x)\stackrel{\neg}\lambda(x)H(x)^2\nonumber\\
  &+&\lambda_E\langle \phi\rangle \lambda(x)\stackrel{\neg}\lambda(x)H(x),
\end{eqnarray}
so we identify a triple coupling between the Elko and the Higgs field, $\alpha_E=\lambda_E\langle \phi\rangle$, where $\langle \phi\rangle=246\gev$. The Elko mass term is modified as follows
\begin{equation}
m^2=m_\varepsilon^2-\lambda_E\frac{\langle \phi\rangle^2}{2}\; .
\label{mass}
\end{equation}

 Many possibilities are opened by this modification in the Elko mass term, e.g., a heavy Elko can lead to more missing energy in the detector, making its identification from background events easier. On the other hand, there is a trade-off with the $\lambda_E$ and $m_\varepsilon$ parameters, which can turn the production cross-section very small. 

To obtain the cross-section we start calculating the Elko spin sum of the $H\stackrel{\neg}{\lambda}^X\lambda^X$  (two Elkos generating a Higgs particle) squared, $X=\left\{A,S\right\}$. The spinor indices will be labeled as $a,b=1,2,3,4$, so  we can  write the amplitude  below
\begin{equation}
\mathcal{M}=\frac{\alpha_E}{m}\lambda^{I,a}_\alpha(\vec{p})\stackrel{\neg}{\lambda}_{\alpha^\prime}^{J,b}(\vec{-p})
\delta^{ab}\, ,
\end{equation}
where $\alpha=\left\{\pm,\mp\right\}$. We stress the fact that, using this definition for the triple coupling vertex, we have a renormalizable expression for this process in the end. We proceed squaring the amplitude $\mathcal{M}$
\begin{equation}
|\mathcal{M}|^2=\frac{\alpha_E^2}{m^2}\lambda^{I,a}_\alpha (\vec{p}) {\lambda^{I,c}_\alpha}^\dagger (\vec{p})
\stackrel{\neg}{\lambda}^{J,b}_{\alpha^\prime}(-\vec{p}){\stackrel{\neg}{\lambda}^{J,d}_{\alpha^\prime}}^\dagger(-\vec{p})
\delta^{ab}\delta^{cd}.
\end{equation}

The unpolarized Elko spin sum can be written  using the relation given in the appendix, as
\begin{eqnarray}
\frac{1}{16}\displaystyle\sum_{\alpha,\alpha^\prime}\displaystyle\sum_{I,J}|\mathcal{M}|^2&=& \frac{\alpha_E}{16m^2}\displaystyle\sum_{\alpha, \alpha^\prime}(\lambda^A_\alpha{\lambda_\alpha^A}^\dagger+\lambda^S_
\alpha{\lambda_\alpha^S}^\dagger)_{ac}\\\nonumber
&\times&\displaystyle\sum_{I}(\stackrel{\neg}{\lambda}_{\alpha^\prime}^I{\stackrel{\neg}{\lambda}_{\alpha^\prime}^I}
^\dagger)_{bd}\delta^{ab}\delta^{cd}\, .
\label{byfrost2}
\end{eqnarray} 

Using  $2(E\mathbb{I} -|\vec{p}|\mawave)_{ac}$  in the first sum of the product above~\cite{elko}:
 \begin{eqnarray} \displaystyle\sum_{\alpha=\left\{-,+\right\},\left\{+,-\right\}}\lambda_\alpha^S(\vec{p})\left(\lambda_\alpha^S(\vec{p})\right)^\dagger=(E-|\vec{p}|)(\mathbb{I}+\mawave)\\ \displaystyle\sum_{\alpha=\left\{-,+\right\},\left\{+,-\right\}}\lambda_\alpha^A(\vec{p})\left(\lambda_\alpha^A(\vec{p})\right)^\dagger=(E+|\vec{p}|)(\mathbb{I}-\mawave),
 \end{eqnarray}
while for  the second sum we can use 
\begin{equation}
\displaystyle\sum_{\alpha^\prime}\stackrel{\neg}{\lambda}_{\alpha^\prime}^I{\stackrel{\neg}{\lambda}_{\alpha^\prime}^I}^\dagger
=\displaystyle\sum_{\alpha}{\lambda_{\alpha^\prime}^I}^\dagger\lambda^I_{\alpha^\prime},
\end{equation} 
thus
\begin{equation}
\frac{1}{16}\displaystyle\sum_{\alpha,\alpha^\prime}\displaystyle\sum_{I,J}|\mathcal{M}|^2=\frac{\alpha^2_E}
{8m^2}\left(E\delta_{bd}-|\vec{p}|\mawave_{bd}\right)\displaystyle\sum_{I}\displaystyle\sum_{\alpha}
{\lambda^{I,b}_\alpha}^\dagger\lambda^{I,d}_\alpha\, .
\label{byfrost3}
\end{equation}

Now making use of Equations (B.24) and (B.25) of reference~\cite{elko}  again, we can write the sum in Equation (\ref{byfrost3}) explicitly
\begin{equation}
\displaystyle\sum_{\alpha}{\lambda_\alpha^I}^\dagger\lambda_\alpha^I=2(E-|\vec{p}|)+2(E+|\vec{p}|)=4E
\end{equation}
so one  can write the unpolarized squared amplitude of the annihilation process as
\begin{eqnarray}
\label{byfrost4}\nonumber
\frac{1}{16}\displaystyle\sum_{\alpha,\alpha^\prime}\displaystyle\sum_{I,J}|\mathcal{M}|^2&=&\frac{\alpha^2_E}
{16m^2}\left[16E^2-2p\displaystyle\sum_{I}\displaystyle\sum_\alpha \left({\lambda^{I,b}_\alpha}^\dagger 
\mawave_{bd}\lambda^{I,d}_\alpha\right)\right]\\
&=&\frac{\alpha^2_E}{16 m^2}\left[16E^2-2|\vec{p}|\, tr\left(2E\mawave +2|\vec{p}|\mathbb{I}\right)\right]=\frac{\alpha^2_E}{m^2}(E^2-p^2)\nonumber\\
&=&\alpha_E^2,
\end{eqnarray} 
in the C.o.M. frame. Since it is energy independent, we are sure that it is renormalizable, maintaining $\alpha_E< 1$ for the convergence of the perturbative series.

The production of dark matter at the LHC is often searched for in conjunction with an accompanying particle that serves as a tagging signature. Processes like $pp\rightarrow \chi+X$, where $\chi$ is a DM candidate and $X$ can be a QCD jet, a photon, or a massive weak boson $W$/$Z$, have been extensively studied at the LHC. The mono-$Z$/$W$ has been proven to be the best channel for the DM search and we explored this process in the case of a light Elko in~\cite{nosso}, where we showed that an $\alpha_E=0.5$ can be probed for an integrated luminosity of $\sim 1$ ab$^{-1}$.

\begin{figure}
\begin{center}
\includegraphics[scale=.4]{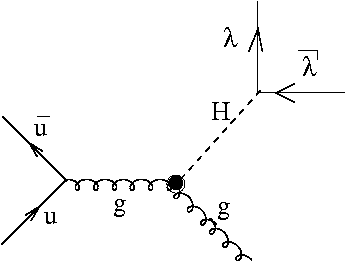}
\end{center}
\caption{A subprocess with a Higgs decay in two Elkos into the monojet production.}\label{tho2}
\end{figure}

Monojet signatures can also be used in the search for DM. To that end, we simulate the process $pp\rightarrow j+\eslash_T$ where the missing transverse energy is due to aa escaping pair of Elkos from the decay of a Higgs boson at the LHC $\sqrt{s}=14$ TeV . A representative Feynman graph is shown in Fig.~(\ref{tho2}).

 The signal events  were generated by modifying  the \verb+Heft+ model present in \verb+MadGraph5+ \cite{mad5} including hadronization with Pythia~\cite{pythia} and detector simulation with PGS~\cite{pgs}.  A mass of $m_H=125.5\gev$ for the Higgs particle was used in the simulation. The Higgs decay width receives an extra contribution from the partial decay width of the decay into two Elkos when compared to the usual parameter included in the program, but this effect is very small and can be safely neglected. 

A light Elko, as it seems to be preferred by relic abundance measurements, leads to little missing energy signatures at the LHC. This makes its separation from background processes much more difficult. For this reason we tried a heavier Elko in order to determine whether we can achieve a bigger signal cut efficiency. The $m$ parameter was adjusted to $50 \gev$, considering  the $m_\varepsilon$ value which fixes $\alpha_E^2$ to $0.17\gev^2$ using its definition from Equation~(\ref{mass}). It is not possible to get much stronger perturbative couplings, but it is possible to have heavier Elkos by adjusting the $m_\varepsilon$ parameter. However, heavier Elkos are produced at lower rates, making its detection harder again. The impact of a heavy Elko on the relics abundance should be investigated further in order to confirm the viability of this scenario. We obtained a cross-section of $27.1$ fb for the signal before cuts.

Concerning the backgrounds, the dominant irreducible one to $j+\eslash_T$ is the process $Z+j\rightarrow j+\nu\bar{\nu}$. We also simulated the two main reducible background processes: $W^\pm+j\rightarrow j+\ell^\pm+\nu$, where the charged lepton is too soft or escapes the fiducial volume of the detector, and QCD $pp\rightarrow jj$, where the missing energy is due to low efficiency in the reconstruction of the jets.

We used \verb+MadGraph5+ again, as in the signal case,  to simulate the backgrounds including hadronization and detector  effects. The following cuts were required both on the signal and on the backgrounds 
\begin{eqnarray} \label{7}\nonumber
p_T(j_1) &>& 150\, \hbox{GeV}, \, p_T(j_2) > 50\, \hbox{GeV}, \, |\eta_j|<5, \, \Delta R (j_1, j_2)>0.4, \label{cuts2}\\
\eslash_T &>&350\, \hbox{GeV} , \, |\Delta\phi(j_1, j_2)-M_Z|<1.2\, \label{cuts1}
\end{eqnarray}

These cuts eliminate all QCD $jj$ events and $Wj$ events as well. On the other hand, the differential distributions of
$Zj$ backgrounds and our signal events are too similar even for a heavy Elko. For the cuts displayed above, about 900 fb cross section for $Zj$ and only a small signal cross section of 0.3 fb are expected. This is a too low signal rate even for the whole amount of data expected at the closure of the LHC, 3 ab$^{-1}$. Even for higher luminosities, an optimal control of systematic uncertainties would be necessary to observe that signal.

Monophoton signatures are expected to be suppressed by a factor $\sim (\alpha_{em}/\alpha_s)^2$ compared to monojets, but the backgrounds are much lower on the other hand. Yet it might be possible to observe missing energy associated to Elkos at the high luminosity regime of the LHC in $\gamma+\eslash_T$ events. Another possibility is the mono-$W$ signature with leptonic or hadronic decays. Exploring the quartic coupling with two Elkos and two Higgs bosons might give us the opportunity to study mono-Higgs signatures with a Higgs boson decaying into bottom quarks, or pairs of massive weak bosons. These are possible studies that we plan to explore in the future.

\section{Conclusions} 
In this work we give an example of a phenomenological study of the Elko field at the CERN LHC. Despite the fact that the signal was very small to constrain the Elko coupling using monojet events at the LHC, we show that  the mass  parameter, now with the contribution of a Higgs vev, can be adjusted to provide some interesting implications for the Elko search at the LHC. As a possible follow-up to this study, we are planning to investigate some further aspects of this theory, including the implication of a heavy Elko for the relics abundance and other collider signatures associated to dark matter searches. Beyond the scope of this paper  we hope that future theoretical studies in the Elko field may include aspects such as the existence of angular asymmetries in physical processes which could give a clear signal in  particle detectors.  

\appendix
\section{}
One of the building blocks of the  Elko spinor  is  its massive Weyl components, $\Phi_L^{+,-}$. So, if we change $\vec{p}\to -\vec{p}$, they transform as
\begin{eqnarray}
\Phi_L^+(\vec{0})&\to& \sqrt{m}\left(\begin{array}{c}-i \sin(\theta/2)e^{-i\phi}/2\nonumber\\
i \cos(\theta/2)e^{i\phi/2}\end{array}\right)=-i\Phi_L^-(\vec{0})\nonumber\\
\Phi_L^-(\vec{0})&\to&\sqrt{m} \left(\begin{array}{c}-i \cos(\theta/2)e^{-i\phi}/2
-i \sin(\theta/2)e^{i\phi/2}\end{array}\right)=i\Phi^+_L(\vec{0}),
\end{eqnarray}
once $\theta\to\pi-\theta$ and  $\phi\to \phi+\pi$, we have $sin(\theta/2)\to cos(\theta/2)$, $cos(\theta/2)\to sin(\theta/2)$, $e^{i\phi/2}\to i e^{i\phi/2}$, and  $e^{-i\phi/2}\to -ie^{-i\phi/2}$.
Using this fact and writing the expressions for the  Elko spinors explicitly, it is easy to see that the following relations hold
\begin{eqnarray}
\lambda^S_{\left\{-,+\right\}}(\vec{0})&=&\left(\begin{array}{c}i\Theta {\Phi_L^+(\vec{0})}^{*} \Phi^+_L(\vec{0})\end{array}\right)\to \left(\begin{array}{c}-\Theta {\Phi_L^-(\vec{0})}^{*} -i\Phi^+_L(\vec{0})\end{array}\right)=-i\lambda^A_{\left\{+,-\right\}}(\vec{0})\nonumber\\
\lambda^A_{\left\{+,-\right\}}(\vec{0})&=&\left(\begin{array}{c}-i\Theta {\Phi_L^-(\vec{0})}^{*} \Phi^-_L(\vec{0})\end{array}\right)\to \left(\begin{array}{c}-\Theta {\Phi_L^+(\vec{0})}^{*} i\Phi^+_L(\vec{0})\end{array}\right)=i\lambda^S_{\left\{-,+\right\}}(\vec{0})\nonumber\\
\lambda^S_{\left\{+,-\right\}}(\vec{0})&=&\left(\begin{array}{c}i\Theta {\Phi_L^-(\vec{0})}^{*} \Phi^-_L(\vec{0})\end{array}\right)\to \left(\begin{array}{c}\Theta {i\Phi_L^+(\vec{0})}^{*} i\Phi^+_L(\vec{0})\end{array}\right)=i\lambda^A_{\left\{-,+\right\}}(\vec{0})\nonumber\\
\lambda^A_{\left\{-,+\right\}}(\vec{0})&=&\left(\begin{array}{c}-i\Theta {\Phi_L^+(\vec{0})}^{*} \Phi^+_L(\vec{0})\end{array}\right)\to \left(\begin{array}{c}\Theta {\Phi_L^-(\vec{0})}^{*} -i\Phi^-_L(\vec{0})\end{array}\right)=-i\lambda^S_{\left\{+,-\right\}}(\vec{0})\, .\nonumber \\
&&
\end{eqnarray}

Therefore $\lambda^{S/A}_{\left\{\pm, \mp\right\}}(\vec{0})=\pm i \lambda^{A/S}_{\left\{\mp, \pm\right\}}(\vec{0})$ when $\phi\to \phi+\pi$ and $\theta\to \pi-\theta$. Since  

\begin{equation}
\lambda^{S/A}_{\left\{\mp, \pm\right\}}(\vec{p})=\sqrt{\frac{E+m}{2m}}\left(1\mp 
\frac{|\vec{p}|}{E+m}\right)\lambda^{S/A}_{\left\{\mp, \pm\right\}}(\vec{0})
\label{byfrost12}
\end{equation} 
so the  Lorentz boost to $\vec{p}$ depends only on $p=|\vec{p}|$, and we can use the fact that $\lambda^{S/A}_{\left\{\pm, \mp\right\}}(-\vec{p})=\pm i \lambda^{A/S}_{\left\{\mp, \pm\right\}}(\vec{p})$. 

We can prove that the same relation holds for a dual Elko spinor
\begin{eqnarray}
\stackrel{\neg}{\lambda}^{S/A}_{\left\{\mp,\pm\right\}}(-\vec{p})&=&\pm i\left[\lambda^{S/A}_{\left\{\pm,\mp\right\}}(-\vec{p})\right]^\dagger\gamma_0=\pm i\left[-i\lambda^{A/S}_{\left\{\mp,\pm\right\}}(\vec{p})\right]^\dagger\gamma_0\nonumber\\
&=&\pm i\stackrel{\neg}{\lambda}^{A/S}_{\left\{\pm,\mp\right\}}\, ,
\end{eqnarray}
or, alternatively
\begin{equation}
\stackrel{\neg}{\lambda}^{S/A}_{\alpha}(-\vec{p})=- i\varepsilon^\beta_\alpha \stackrel{\neg}{\lambda}^{A/S}_{\beta}(\vec{p}).
\end{equation}
where $\varepsilon^{\left\{-,+\right\}}_{\left\{+,-\right\}}=-1=-\varepsilon^{\left\{+,-\right\}}_{\left\{-,+\right\}}$.

Finally, we can write
\begin{eqnarray}
\displaystyle\sum_{I=S,A}\displaystyle\sum_{\alpha}\stackrel{\neg}{\lambda}^I_\alpha(-\vec{p}){\stackrel{\neg}{\lambda}^I_\alpha(-\vec{p})}^\dagger&=&\displaystyle\sum_{J=S,A}\displaystyle\sum_{\beta}\left[-i\varepsilon^\beta_\alpha\stackrel{\neg}{\lambda}^J_\alpha(\vec{p})(+i)\varepsilon^\beta_\alpha{\stackrel{\neg}{\lambda}^J_\beta(\vec{p})}^\dagger\right]\nonumber\\
&=&\displaystyle\sum_{J=S,A}\displaystyle\sum_{\beta}\left[\stackrel{\neg}{\lambda}^I_\alpha(\vec{p}){\stackrel{\neg}{\lambda}^I_\beta(\vec{p})}^\dagger\right]\, ,
\end{eqnarray}
and the standard relations for the spin sums, with all elements with the same momentum,  can be used.

\end{document}